\documentclass{tMOP2eEmpty}

 \newcommand{\vek}    [1] {\textrm{\textbf{{#1}}}} 
 \newcommand{\vekk}   [1] {{\bm  #1} } 
 \newcommand{\Op}     [1] {\hat{\cal{#1}}} 

 \newcommand{\ket}    [1] {| #1 \rangle}

\begin{document}

\title{Ionisation of H$_2$ in intense ultrashort laser pulses: parallel
         versus perpendicular orientation}

\author{Yulian V.~Vanne and Alejandro Saenz\\\vspace{6pt}
        AG Moderne Optik, Institut f\"ur Physik, Humboldt-Universit\"at
        zu Berlin,\\
        Hausvogteiplatz 5-7, D\,--\,10\,117 Berlin, Germany}

\maketitle
\begin{abstract}
A theoretical comparison of the electronic excitation and ionisation
behaviour of molecular hydrogen oriented either parallel or perpendicular
to a linear polarised laser pulse is performed. The investigation is
based on a non-perturbative treatment that solves the full time-dependent
Schr\"odinger equation of both correlated electrons 
within the fixed-nuclei approximation and the dipole
approximation. Results are shown for two different laser pulse lengths and
intensities as well as for a large variety of photon frequencies starting
in the 1- and reaching into the 6-photon regime. In order to investigate
the influence of the intrinsic diatomic two-center problem even further,
two values of the internuclear separation and a newly developed atomic
model are considered.
\end{abstract}
\bigskip

\bigskip

\section{Introduction}
Despite the numerous experimental work on molecules in intense ultrashort 
laser pulses (see, e.\,g., \cite{sfm:post04} for a review) the theoretical 
treatment of even the simplest neutral molecule, H$_2$, remains a big 
challenge. In view of the interesting molecular strong-field phenomena like 
bond softening, bond hardening, or enhanced ionisation a deeper theoretical 
understanding of the molecular behaviour is desirable. This interest is 
further stimulated by recent progress in the direction of time-resolved 
molecular orbital tomography \cite{sfm:itat04} or visualisation of nuclear 
dynamics with sub-femtosecond 
resolution \cite{sfm:niik02,sfm:niik03,sfm:bake06,sfm:goll06,sfm:ergl06}. 
A key problem is the understanding of the relation of 
molecular structure on the strong-field response. There is, of course, 
the nuclear rotational and vibrational degrees of freedom them self which 
should be considered. Furthermore, even the purely electronic response 
of a molecule differs from the atomic one, since the electron density 
is anisotropic and depends on the nuclear geometry. 

Only recently {\it ab initio} calculations for H$_2$ exposed to strong 
fields became available in which both electrons are treated in full 
dimensionality. This includes calculations of the ionisation and 
dissociation behaviour of H$_2$ exposed to an intense static or 
quasi-static electric field
\cite{sfm:saen00a,sfm:saen00b,sfm:saen02a,
sfm:saen02b,sfm:saen02c,sfm:saen04} 
which revealed the possible occurrence of bond softening and enhanced 
ionisation in {\it neutral} H$_2$ due to a field-induced avoided 
crossing of the neutral ground state with the ion-pair state H$^+$H$^-$. 
Shortly thereafter, a full time-dependent calculation confirmed 
this finding \cite{sfm:haru00,sfm:haru02,sfm:kono06}. These calculations  
were based on a judiciously 
chosen grid on which the electronic wavefunctions were expanded. 
In a different approach the electronic wave-packet is expanded in 
terms of field-free eigenstates of H$_2$ \cite{sfm:awas05,sfm:awas06}. 
The same {\it ansatz}, but using a one-centre expansion for the 
electronic orbitals was used in \cite{sfm:pala06}. In that work even 
vibrational motion was included within the Born-Oppenheimer 
approximation. All these works did, however, only consider 
a parallel orientation of the linear polarised laser field 
and the molecular axis. The reason is rather simple: in this case 
the problem of 6 spatial dimensions describing two electrons moving 
in the field of 
the laser and the two nuclei reduces effectively to a 5-dimensional 
one, because the quantum number of angular momentum along the 
molecular axis, $M$, is conserved or, equivalently, the cylindrical 
symmetry of H$_2$ is not broken by the electromagnetic field.  

Correspondingly, the consideration of the orientational dependence 
of the strong-field behaviour of H$_2$ is rather limited so far. 
Within lowest-order perturbation theory (LOPT) a comparison of the 
ionisation rates for parallel and perpendicular orientation were 
presented for 2- and 3-photon ionisation and some internuclear 
separations in \cite{sfm:apal02}. More recently, the orientational 
dependence of H$_2$ was also considered in full time-dependent 
treatments using either time-dependent density-functional theory (TD-DFT) 
\cite{sfm:uhlm06} or a single-active electron approximation (SAE) 
\cite{sfm:niko07}. The validity of the SAE (and simplified models 
like the molecular Ammosov-Delone-Krainov tunnelling model (MO-ADK) 
or the molecular strong-field approximation (MO-SFA)) that reduces 
the problem to three spatial dimensions was recently 
investigated in \cite{sfm:awas08}. There it was found that 
especially in the few-photon case its applicability is quite 
restricted, since there is a pronounced dependence on the photon 
frequency due to the possible occurrence of resonantly enhanced 
multi-photon ionisation. Since the position of those resonances 
depends strongly on the excitation energies, a simplified model 
like SAE can lead to wrong positions of the resonances and thus 
a very wrong prediction for the ionisation rate at a given 
photon frequency. 

In the present work the previously developed approach \cite{sfm:awas05} 
for the full-dimensional {\it ab initio} treatment of the electronic 
motion in H$_2$ exposed to an intense laser pulse was extended to the 
consideration of a perpendicular orientation of the molecular axis with 
respect to the field. This means that now all 6 spatial dimensions 
of the two correlated electrons are explicitly considerd. 
The ionisation and electronic excitation 
is compared for parallel and perpendicular orientation as a function 
of photon frequency, completely covering the 2- to 5-photon regime 
(and partly extending into the 1- and 6-photon regimes). Furthermore, 
two different laser-pulse lengths and intensities as well as two 
internuclear separations are considered.  

For some time it was believed that from an experimental point of view 
the parallel orientation of H$_2$ is the most important one, since 
an intense laser pulse will align the molecule (due to the difference 
in polarisability parallel and perpendicular to the molecular axis) 
\cite{sfm:char94,sfm:post98,sfm:post98a,sfm:post04}. 
For a sufficiently long laser pulse this alignment will take place 
during the raising edge of the laser pulse at intensities smaller 
than the peak intensity at which ionisation occurs. Therefore, 
the ionisation signal will reflect the ionisation 
behaviour of a parallel aligned H$_2$ molecule. However, for very 
short pulses that are nowadays available 
even H$_2$ will not have the time to align to the laser field. 
In such a case a comparison to experimental data requires the 
calculation of the ionisation yield of an isotropic or partly 
aligned sample. Furthermore, short laser pulses with peak 
intensities too small to cause substantial ionisation are in turn 
used to generate rotational wave-packets \cite{sfm:lars99b,sfm:stap03}. 
If an ionising laser pulse 
follows such a pre-pulse with a well-defined time delay, it is 
possible to control the degree of alignment or even anti-alignment, 
i.\,e.\ to investigate the strong-field behaviour as a function of 
the expectation value of the angle between the molecular axis and 
the laser field. This was done for the investigation of the orientational 
dependence of high-harmonic generation \cite{sfm:velo01} 
or ionisation, e.\,g., for molecular 
nitrogen \cite{sfm:litv03,sfm:zeid06}. The latter results motivated 
corresponding theoretical studies within single-active-electron based 
strong-field approximations which showed disagreeing and gauge-dependent 
results for the orientational dependence of the ionisation rate 
\cite{sfm:kjel04,sfm:beck04,sfm:kjel06,sfm:usac06}. 
Therefore, the investigation of the orientational dependence of the 
strong-field behaviour based on a full solution of the time-dependent 
Schr\"odinger equation is very timely.

\section{Method}
The full-dimensional time-dependent Schr\"{o}dinger equation 
(TDSE) is solved by expanding the time-dependent wave function 
in terms of field-free states. The latter are obtained from a 
configuration-interaction (CI) calculation in which the 
Slater determinants are formed with the aid of H$_2^+$
wave functions expressed in terms of $B$ splines in prolate 
spheroidal coordinates ($1\leq \xi < \infty, -1\leq \eta\leq 1, 
0\leq \phi < 2\pi$). The electronic structure CI method is 
discussed in detail in \cite{dia:vann04}. Note, electron-electron 
interaction is not included in the orbitals used for the CI calculation. 
Such an approach is especially suitable for asymmetric excited states
(for example a state where one electron is left in the lowest 
lying orbital while the other one is ionised), but is not 
perfect for describing, e.\,g., the (symmetric) electronic 
ground state. Nevertheless, it has been shown in \cite{dia:vann04} 
that with the present approach very accurate 
ground-state energies can be obtained for H$_2$, if the 
basis set is chosen judiciously. In calculations as they are 
discussed here, a compromise has to be searched for, 
since the goal is to achieve a rather uniform accuracy 
with respect to the description of a plethora of states,
from the ground state up to very energetic continuum ones. 

A key feature of the present approach is the discretisation of the 
electronic continuum. This is a consequence of the chosen $B$-spline basis 
confined within a finite spatial volume defined by $\xi_{\rm max}$. 
Therefore, only states that are confined within this volume or 
that have a node at the volume boundary are obtained 
(see \cite{dia:vann04} for a more 
detailed discussion). Since the code allows to calculate the 
electronic states of any symmetry (singlet, triplet, $\Sigma$, 
$\Pi$, etc.) it is also possible to perform calculations for any 
possible orientation of the molecular axis with respect to the 
polarisation of the laser field. The use of the prolate spheroidal 
coordinate system allows to solve the electronic problem for arbitrary 
values of the internuclear distance \cite{dia:vann06}. This is an 
important advantage over the one-centre approximation. The latter was 
extensively used for H$_2$ in theoretical single-photon ionisation 
studies by F.~Mart{\'\i}n and collaborators (for a review see 
\cite{bsp:mart99}). Also previous multi-photon studies on H$_2$ within 
LOPT adopted that approach \cite{sfm:apal02,sfm:pala06}. However, 
the one-centre expansion converges very slowly for large internuclear 
distances. 

The solution of the TDSE describing molecular hydrogen exposed 
to a laser field follows closely 
the approach that has successfully been used for one- and 
two-electron atoms before (see \cite{sfa:lamb98a} for a review).  
The total in-field Hamiltonian is given by%
\begin{equation}
         \Op{H} = \Op{H}_0 + \Op{V}(t)
\end{equation}
where $\Op{H}_0$ is the field-free electronic Born-Oppenheimer 
Hamiltonian of a hydrogen molecule and $\Op{V}(t)$ is the 
operator describing its interaction with the (time-dependent) 
laser field. The non-relativistic approximation is used for both 
operators, and the interaction with the laser field
is described within the dipole approximation and in velocity gauge.
For a linearly polarised laser field with the polarisation axis
$\vekk{\epsilon}$ the interaction operation is given by%
\begin{equation}
      \Op{V}(t) = - A(t) \vekk{\epsilon}\cdot\vek{P}
\end{equation}
(Here and in the following atomic units ($e=m_e=\hbar=1$) are used 
unless specified otherwise.) $A(t)$ is the magnitude of the vector 
potential 
of the laser field and $\vek{P}$ is the total momentum operator of 
the electrons.

The present work is restricted to the case where the  
H$_2$ molecule before the pulse is in its ground $1\Sigma_g^+$ state.
For parallel orientation, $\vekk{\epsilon}\,\, || \,\, \vek{R}$, only
transitions from $\Sigma_g^+$ to $\Sigma_u^+$ and vice versa 
are allowed. Therefore,
only two symmetries have to be considered in this case. For
perpendicular orientation, $\vekk{\epsilon} \perp \vek{R}$, the transitions
$ \Sigma_g^+ \leftrightarrow \Pi_u \leftrightarrow 
    \Delta_g \leftrightarrow \Phi_u \dots $
are allowed. Moreover, all states with the symmetries $\Pi$,  $\Delta$, \dots,
i.\,e.\ with the absolute value of the component the total angular momentum
along the internuclear axis $\Lambda > 0$, are double degenerate, since
one has for the value of the total angular momentum
along the internuclear axis $M = \pm \Lambda$. However, the explicit 
use of the reflection symmetry (here and in the following with reflection 
the reflection plane containing the molecular axis is meant) helps to
reduce the dimensionality of the problem. Indeed, both $\Op{H}_0$ and 
$\Op{V}(t)$ are symmetric with respect to the reflection operation and 
the same is true for the initial $1\Sigma_g^+$ state. Therefore, only the 
linear combinations of two degenerate states which are symmetric with 
respect to the reflection transformation have to be considered. 
(If the initial state would be $1\Sigma_g^-$, the linear combinations had 
to be antisymmetric.) 

The resulting TDSE%
\begin{equation}\label{tdse}
    i  \,\frac{\partial \ket{\Psi} }{\partial t}  \;=\; \Op{H} \ket{\Psi}  
\end{equation}
is solved by expanding the wave function $\ket{\Psi}$ according to
\begin{equation}\label{expansion}
     \ket{\Psi (t)} \;=\; \sum_{n\Omega} \: C_{n\Omega}(t)\, \ket{\phi_{n \Omega}}
\end{equation}
in terms of the time-independent wave functions $\ket{\phi_{n \Omega}}$. The 
latter are solutions of the field-free molecular Schr\"odinger equation%
\begin{equation}\label{H0}
       \Op{H}_0 \ket{\phi_{n \Omega}} 
                   \;=\; E_{n \Omega}\: \ket{\phi_{n \Omega}} \quad .
\end{equation}
The two-electron wavefunctions $\ket{\phi_{n \Omega}}$ are orthonormal 
and symmetric with respect to the reflection symmetry.
The compound index $\Omega$ represents $\Lambda$ and the parity with
respect to inversion symmetry ({\it gerade} or {\it ungerade}). 
The $n$ is just an index of a state with a particular 
symmetry $\Omega$. Due to the chosen approach for 
solving Eq.\,(\ref{H0}) (using a CI expansion based on H$_2^+$ orbitals 
that are expanded in a $B$-spline basis contained in a finite box) all 
states are discretised as was mentioned before. Therefore, the index $n$ 
remains discrete even for states in the electronic continuum. 
In the case of perpendicular orientation the summation in 
Eq.\,(\ref{expansion}) is restricted to $\Lambda \le \Lambda_{\rm max}$.

Since the CI method \cite{dia:vann04} generates only solutions 
$\psi_{n, \Omega}$ which have $M = \Lambda$, they must be adapted for
the present purpose. Although for $\Lambda = 0$ they are equivalent to
$\phi_{n \Omega}$, for $\Lambda > 0$ the following linear combination
has to be used
\begin{equation}\label{phi_nOmega}
   \phi_{n \Omega} \;=\;
        ( \psi_{n, \Omega} +  \psi_{n, \Omega}^* )/\sqrt{2} \quad.
\end{equation}
As discussed in \cite{dia:vann04}, with a proper normalisation of 
$\psi_{n, \Omega}$ the reflection transformation is equivalent to a 
complex conjugation of the wavefunction. Therefore, the definition
(\ref{phi_nOmega}) ensures that $\phi_{n \Omega}$ is symmetric with respect
to reflection.

Substitution of Eq.\,(\ref{expansion}) into the TDSE [Eq.\,(\ref{tdse})], 
multiplication of the result by $\phi_{n'\Omega'}^*$, and integration over the 
electronic coordinates yields%
\begin{equation}\label{coeff}
 i  \frac{\partial}{\partial t} C_{n'\Omega'} (t) \;=\; 
     E_{n'\Omega'} C_{n'\Omega'} (t) \:  + 
    i A(t) \: \sum_{n \Omega} \: D_{n' \Omega',n \Omega} \: C_{n \Omega}(t) 
\end{equation}
with $D_{n' \Omega',n \Omega}  = \langle \phi_{n' \Omega'} \lvert  
\vekk{\epsilon}\cdot (\nabla_1 + \nabla_2) \rvert \phi_{n \Omega} \rangle$.
As can be shown using Eq.\,(\ref{phi_nOmega}), 
these matrices are related to those obtained in \cite{dia:vann04},
$\bar{D}_{n' \Omega',n \Omega}  = \langle \psi_{n' \Omega'} \lvert  
\vekk{\epsilon}\cdot (\nabla_1 + \nabla_2) \rvert \psi_{n \Omega} \rangle$,
by the relation%
\begin{equation}\label{relation}
D_{n' \Omega',n \Omega}  = \left\{
\begin{array}{rl}
\displaystyle \sqrt{2}\bar{D}_{n' \Omega',n \Omega} & 
                        \quad \mbox{for $\Lambda + \Lambda'=1$} \\
\displaystyle  \bar{D}_{n' \Omega',n \Omega} &
                        \quad \mbox{otherwise}. 
\end{array} \right.
\end{equation}
Here, the reality of $\bar{D}_{n' \Omega',n \Omega}$ and
the identity for $\Lambda >0$,%
\begin{equation}
\langle \psi_{n' \Omega'} \lvert \vekk{\epsilon}\cdot (\nabla_1 + \nabla_2) 
\rvert \psi^*_{n \Omega} \rangle = 0, \quad 
\mbox{if $\Lambda' \neq 0$ or $\Lambda \neq 1$}  \quad ,
\end{equation}
was used.

It should be emphasised that with this approach the 
complete time dependence is incorporated in the coefficients $C_{n \Omega}$. 
They are calculated by propagating Eq.\,(\ref{coeff}) numerically in 
time using a variable-order, variable-step Adams solver for ordinary 
first-order differential equations. The laser-pulse parameters are 
contained in $A(t)$. Different choices for the temporal shape 
of the pulse are implemented, but in this work only results  
for $\cos^2$-shaped pulses are shown.

For the considered laser parameters converged results were obtained 
using along the 
$\xi$ coordinate 350 $B$ splines of order $k=15$ with a linear knot 
sequence. A box size (defined implicitly by $\xi_{\rm max}$ that 
depends on $R$) of about 350\,$a_0$ is chosen for
both $R = 1.4\,a_0$ and  $R = 2.0\,a_0$.  Along the $\eta$ coordinate 
30 $B$ splines of order 8 were used in the complete 
interval $-1\,\leq \eta \leq +1$, but using the symmetry of a 
homonuclear system as is described in \cite{dia:vann04}.
Out of the resulting 5235 orbitals for every symmetry only 3490 orbitals
were further used to construct configurations, whereas those orbitals 
with highly oscillating angular part (with more than 19 nodes for the 
$\eta$-dependent component) were omitted.
In most of the subsequent CI calculations approximately 6000 
configurations were used for every symmetry. 
These states result from very long configuration series (3490 
configurations) in which one electron occupies the H$_2^+$ ground-state 
$1\,\sigma_g$ orbital while the other one is occupying one of the 
remaining, e.\,g.,  
$n\,\pi_u$ or $n\,\delta_g$ orbitals. The other configurations 
represent doubly excited situations and are responsible for 
describing correlation (and real doubly excited states).
To speed up the TDSE calculations an energy cut-off was used as an additional 
parameter.
Only CI states with an energy below this cut-off (in the present case set 
to about 
300\,eV) were included in the time propagation. For perpendicular orientation
$\Lambda_{\rm max} = 3$ (the final number of states used in
TDSE calculations is about 21,500) gives sufficient convergence 
for total yields, as is checked by a comparison to analogous 
calculations with $\Lambda_{\rm max} = 5$ (about 32,000 states).

\section{Atomic model}
\label{sec:modelatom}
A molecule treated in the fixed-nuclei approximation differs from an 
atom due to the anisotropy of the electronic charge distribution which 
occurs even for the totally symmetric ground state. Alternatively, 
this anisotropic charge distribution may be described in the language of the 
linear combination of atomic orbitals (LCAO) as a multi-centre structure 
that can give rise to interference phenomena. For the analysis of the 
effects of the anisotropy and thus the corresponding orientational 
dependence it is therefore of interest to compare the molecular results 
with the ones obtained for an artificial atom with an isotropic, 
single-centred charge distribution. Since strong-field ionisation is 
known to be very sensitive to the electronic binding energy and the 
exact form of the long-ranged Coulomb potential, it is important that 
the artificial atom agrees in these properties with the molecule. For 
this purpose the simple one-parameter model potential
\begin{equation}
\label{eq:Vatom}
  V(r) = -\frac{1}{r} \left\{ 1 + \frac{\alpha}{|\alpha|} 
          \exp\left[- \frac{2r}{|\alpha|^{1/2}}\right] \right\}
\end{equation}
may be introduced. It satisfies $V(r)\rightarrow - 1/r$ for 
$r\rightarrow \infty$ and reduces to the potential of atomic hydrogen,
$I_p({\rm H})= 0.5\,$a.\,u.,  
for $\alpha \rightarrow 0$. Although the exact ionisation potential $I_p$ 
for arbitrary values of the parameter $\alpha$ can be obtained only 
numerically, it can be quite well approximated by the following expression%
\begin{equation}
\label{eq:IpApprox}
   I_p(\alpha) \approx
 I_p({\rm H}) + \frac{\alpha}{(1+\sqrt{|\alpha|})^s}
\end{equation}
where $s=1$ for $\alpha>0$ and $s=11/4$ for $\alpha<0$. 
For $|\alpha| \ll 1$ the ionisation potential is simply given by 
$I_p(\alpha) \approx I_p({\rm H}) + \alpha$.
\begin{figure}
\begin{center}
\includegraphics[width=8.0cm]{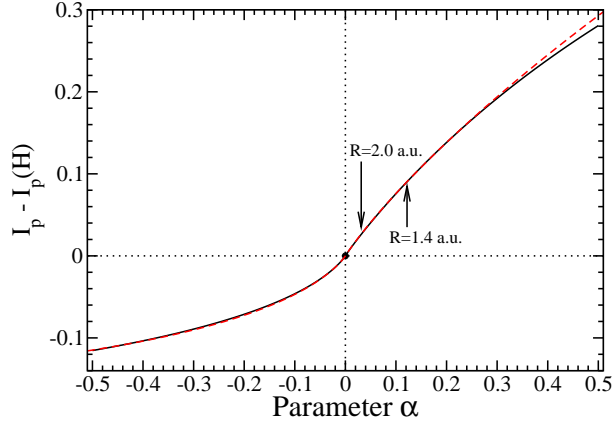}%
\caption{\label{fig:Atom} (Color online) 
Deviation (shift) of the ionisation potential $I_p$ of the atomic-model 
potential (\ref{eq:Vatom})  
from its value for $\alpha=0$, $I_p({\rm H}) = 0.5\,$a.\,u., 
as a function of parameter $\alpha$. The dashed line shows the approximation
given by the expression (\ref{eq:IpApprox}). The arrows indicate the values
of $\alpha$ used in the present work.} 
\end{center}
\end{figure}

Fig.\,\ref{fig:Atom} shows the variation of the ionisation potential 
of the model atom as a function of $\alpha$. More precisely, it shows its 
deviation from the value of the hydrogen atom ($I_p({\rm H}) = 0.5\,$a.\,u.). 
The values of $\alpha$ used in this work were 0.12194 and 0.03126 
(indicated by the arrows in Fig.\,\ref{fig:Atom}) for 
simulating an artificial atom with the same ionisation potentials 
($0.59037\,$a.\,u.\ and $0.52615\,$a.\,u.) as the computed ionisation 
potentials of H$_2$ for the internuclear separation of $R=1.4\,a_0$ 
and $R=2.0\,a_0$, respectively. The dashed line on Fig.\,\ref{fig:Atom}
shows approximate ionisation potential given by 
the expression (\ref{eq:IpApprox}), which yields, respectively,
$0.59038\,$a.\,u.\ and $0.52656\,$a.\,u.\ for the values of 
$\alpha$ given above.

It should be noted that the model 
atom is a one-electron (hydrogen-like) atom in which the effect of both 
the anisotropy due to the two nuclei and due to the second electron in 
H$_2$ is solely contained as a screening of the Coulomb potential modifying 
the ionisation potential. Therefore, this model does not describe any 
excitation or relaxation of a second electron that can occur in the H$_2$ 
calculation. In order to compare to H$_2$ the atomic results were 
multiplied by a factor 2 in order to account for the two equivalent 
electrons in H$_2$. This procedure for comparing SAE results with 
full two-electron calculations were shown to be reasonable (for not 
too high ionisation yields exceeding about 10\,\%) in \cite{sfm:awas08}.

\section{Results}
\begin{table}
\tbl{Electronic energies $E$ (in a.\,u.) of various H$_2$ states as they 
are obtained with the basis sets used in this work and the 
resulting resonant $N$-photon transition frequencies $\omega$ (in eV) and 
wavelengths $\lambda$ (in nm). The last row shows the ground-state energy 
of H$_2^+$ and corresponding 1-photon ionisation threshold.}
{\begin{tabular}{@{}lcccrrrr}\toprule
State & $E^{\rm a}$, a.u. & $E^{\rm b}$, a.u. & 
$N$ & $\omega^{\rm a}$, eV & $\omega^{\rm b}$, eV &
$\lambda^{\rm a}$, nm & $\lambda^{\rm b}$, nm \\
\colrule
1 ${}^1\Sigma_g^+$ (X) & -1.160351 & -1.128787 & 
&  &  & & \\
2 ${}^1\Sigma_g^+$ (EF)& -0.690087 & -0.716303 & 
2 & 6.3982  & 5.6121 & 193.778 & 220.922 \\
&&&
4 & 3.1991  & 2.8060 & 387.556 & 441.844 \\
3 ${}^1\Sigma_g^+$ (GK) & -0.626453 & -0.660305 & 
2 & 7.2640  & 6.3739 & 170.682 & 194.515 \\
&&&
4 & 3.6320  & 3.1870  & 341.364 & 389.030 \\
1 ${}^1\Sigma_u^+$ (B)& -0.702364 & -0.745749 & 
1 & 12.4623 & 10.4229 & 99.486  & 118.953 \\
&&&
3 & 4.1541 & 3.4743   & 298.458 & 356.859 \\
2 ${}^1\Sigma_u^+$ (B')& -0.627569 & -0.663476 & 
1 & 14.4975 & 12.6616  & 85.520 & 97.920 \\
1  ${}^1\Pi_u$ (C)     & -0.687338 & -0.716903 & 
1 & 12.8712 & 11.2078  & 96.326 & 110.622 \\
&&&
3 & 4.2904  & 3.7359  & 288.978 & 331.866 \\
2 ${}^1\Pi_u$ (D)     & -0.623117 & -0.654839 & 
1 & 14.6187 & 12.8966 & 84.811 & 96.136 \\
1 ${}^1\Delta_g$ (J)  & -0.625213 & -0.657517 & 
2 & 7.2808 & 6.4119  & 170.286 & 193.364 \\
2 ${}^1\Delta_g$ (S)  & -0.601098 & -0.633603 & 
2 & 7.6089 & 6.7372 & 162.943 & 184.026 \\
&&&&&&&\\
1 $\sigma_g$ [H$_2^+$]  & -0.569984 & -0.602634 & 
1 & 16.0645 & 14.3171 & 77.178 & 86.597 \\
\botrule
\end{tabular}}
\tabnote{$^{\rm a}$ For the internuclear distance $R=1.4\,a_0$}%
\tabnote{$^{\rm b}$ For the internuclear distance $R=2.0\,a_0$}%
\label{tab:data}
\end{table}
\begin{figure}
\begin{center}
\includegraphics[width=12.0cm]{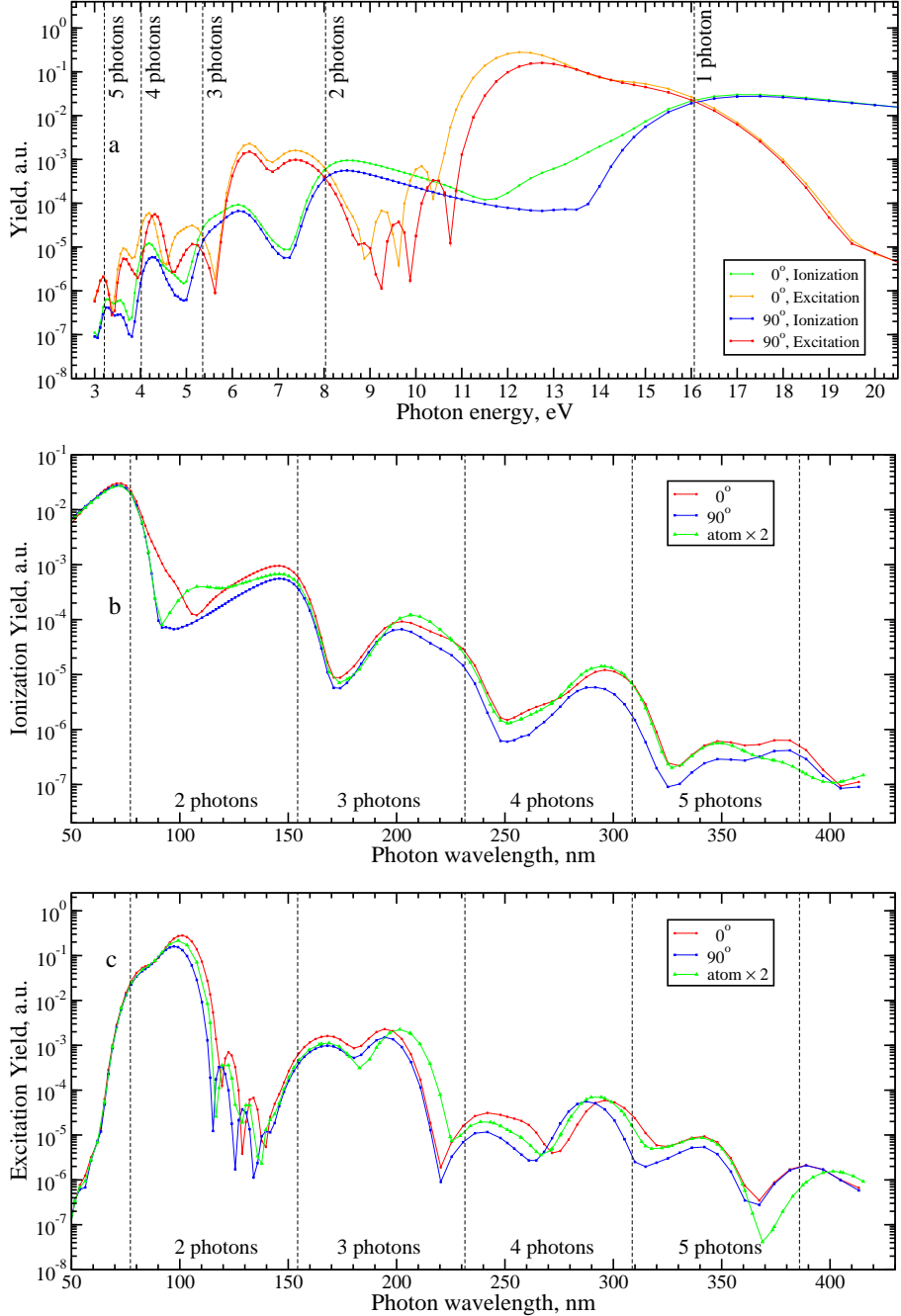}%
\caption{\label{fig:10R1p4} (Color online) 
Ionisation and excitation yields of an 
H$_2$ molecule with fixed internuclear distance $R = 1.4\,a_0$  
for a 10-cycle linear-polarised laser pulse with 
peak intensity $10^{13}$\,W/cm$^2$ and either parallel or perpendicular 
orientation of the molecular axis with respect to the field. Ionisation 
and bound-state excitation yields are shown together as a function 
of the photon energy in eV (a), while b (c) shows 
the ionisation (excitation) yield as a function of photon wavelength. 
In the latter cases, also the yields obtained for a model atom  
(multiplied by a factor 2, see Sec.\,\ref{sec:modelatom}) are plotted. 
The $N$-photon thresholds (for $N=1$ to 5) are indicated by vertical 
dashed lines.} 
\end{center}
\end{figure}
\begin{figure}
\begin{center}
\includegraphics[width=12.0cm]{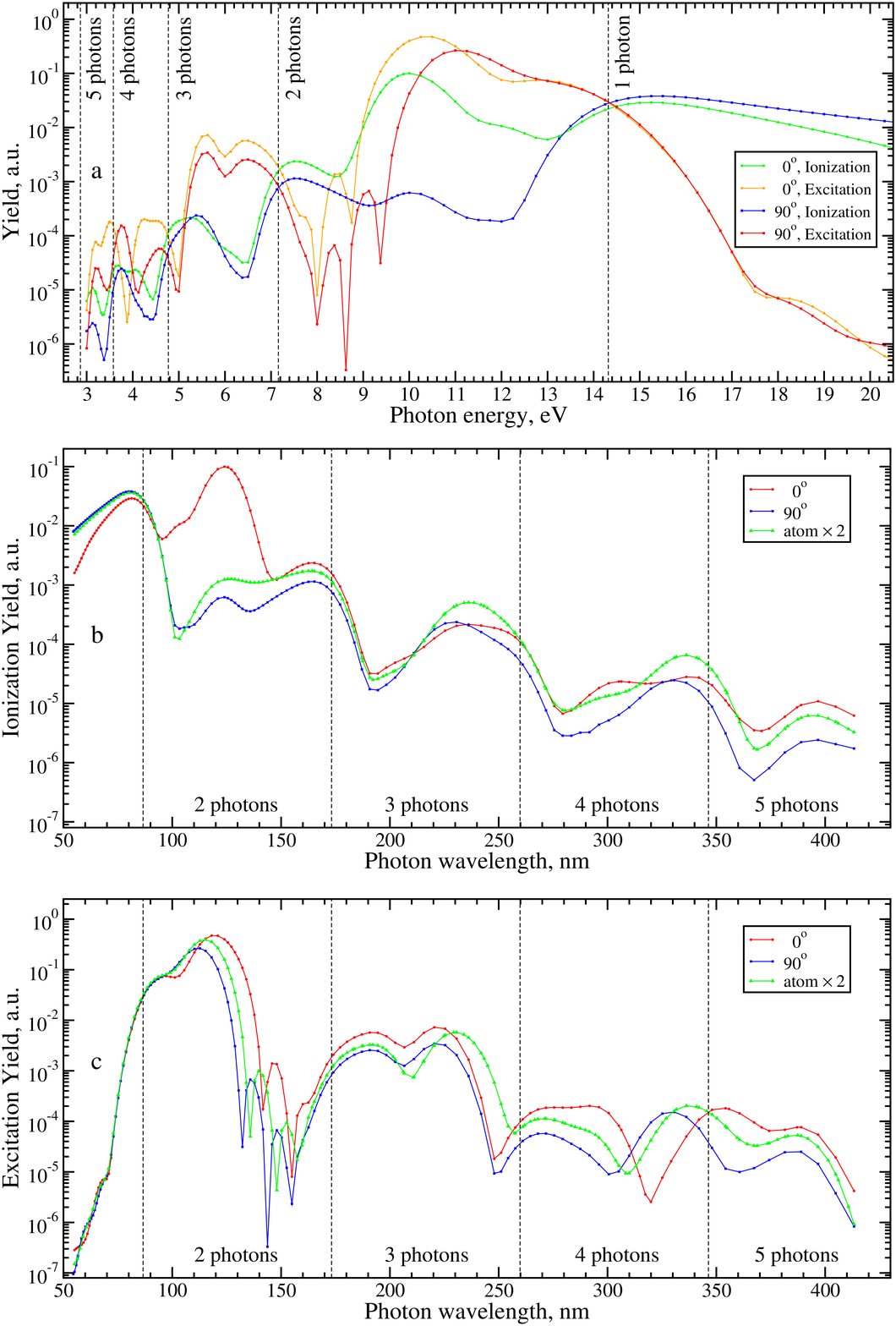}%
\caption{\label{fig:10R2p0} (Color online) As Fig.\,\ref{fig:10R1p4}, but for 
internuclear distance $R = 2.0\,a_0$.} 
\end{center}
\end{figure}
For the subsequent discussion it is helpful to keep in mind the 
relevant energies and transition frequencies (or wavelengths) 
of a number of electronic bound states of H$_2$ that can resonantly 
be excited by a laser with the corresponding photon frequency. 
Since the exact positions of the resonances depend on the adopted 
electronic structure model, Table \ref{tab:data} reports the 
energies obtained with the present approach and basis set. As 
was discussed before, the basis set was chosen to provide a good 
compromise for describing a large number of states and can, of course, 
not compete with a high-precision calculation optimised for a 
single electronic state. Furthermore, Table \ref{tab:data} provides 
the ground-state energy of H$_2^+$ which allows to calculate the 
exact position of the different $N$-photon thresholds. (The 
1-photon threshold is given explicitly.)

Fig.\,\ref{fig:10R1p4} shows the ionisation and electronic excitation 
yields for H$_2$ exposed to laser pulses with a total duration of 
10 cycles, a peak intensity $I=10^{13}$\,W/cm$^2$, and a variable 
photon energy obtained within the fixed-nuclei approximation for 
the nuclear separation $R=1.40\,a_0$ that corresponds 
(to a good approximation) to the equilibrium distance of the field-free 
H$_2$ molecule. The excitation yield $Y_{\rm exc}$ is defined as the  
population of all possible electronically bound excited states, 
i.\,e.\ $Y_{\rm exc} = 1 - P_{\rm gs} - Y_{\rm ion}$ where $P_{\rm gs}$ 
is the population left in the electronic ground state and $Y_{\rm ion}$ 
is the ionisation yield. (Populations and yields are defined in such a way 
that a value of 1 corresponds to 100\,\%.) 
The results obtained for a parallel orientation of the molecular axis  
with respect to the field that are shown in the upper graph agree 
qualitatively to the ones 
obtained for different laser parameters (peak intensities were 
either $2\times 10^{12}$ or $2\times 10^{14}\,$\,W/cm$^2$ and the 
pulse duration was fixed to 15\,fs) in an earlier 
work \cite{sfm:awas05}. However, one notices that the present results 
are much less structured, despite the larger number of data points 
and correspondingly higher resolution. Furthermore, the ionisation 
thresholds marking the transition from an $N$- to an $(N-1)$-photon 
process are by far not as sharp as in \cite{sfm:awas05}. This is 
especially evident for the threshold dividing the 1- and 2-photon 
ionisation regimes. The reason is the (especially for those large photon 
energies) much shorter pulse duration of the present 10-cycle 
pulse compared to a 15\,fs pulse used in \cite{sfm:awas05}. 

A short pulse duration leads to a broad spectral width of 
the Fourier-limited pulse. Fixing the number of cycles instead of 
the total pulse duration leads, of course, to a variation of the 
spectral width as a function of the photon energy. On the other hand, 
a fixed number of cycles has the advantage that one expects the pulses 
for different photon energy to become better comparable with respect 
to adiabaticity of the process, since a fixed pulse duration can lead 
in an extreme case to pulses comprising of in one case many and in the 
other case even less than a single cycle. An effect of the shorter 
pulse duration can, for example, be seen from the (1+1)-REMPI peaks 
that are caused by resonant one-photon transitions to the B and the B' 
$^1\Sigma_u$ states (cf.\ Table \ref{tab:data}). They are clearly 
visible in both the excitation 
and ionisation yields in \cite{sfm:awas05}, but appear in 
Fig.\,\ref{fig:10R1p4}\,a as an almost structureless broad peak in the 
excitation yield spanning the photon energy range from about 11 to 16\,eV. 
Another remarkable difference to the earlier result obtained for 
fixed (longer) pulse length is the fact that in this  
energy window the resonantly enhanced ionisation yield increases 
almost uniformly with photon energy while in the earlier result 
there was a pronounced peak at the position of the B state. 

\begin{figure}
\begin{center}
\includegraphics[width=12.0cm]{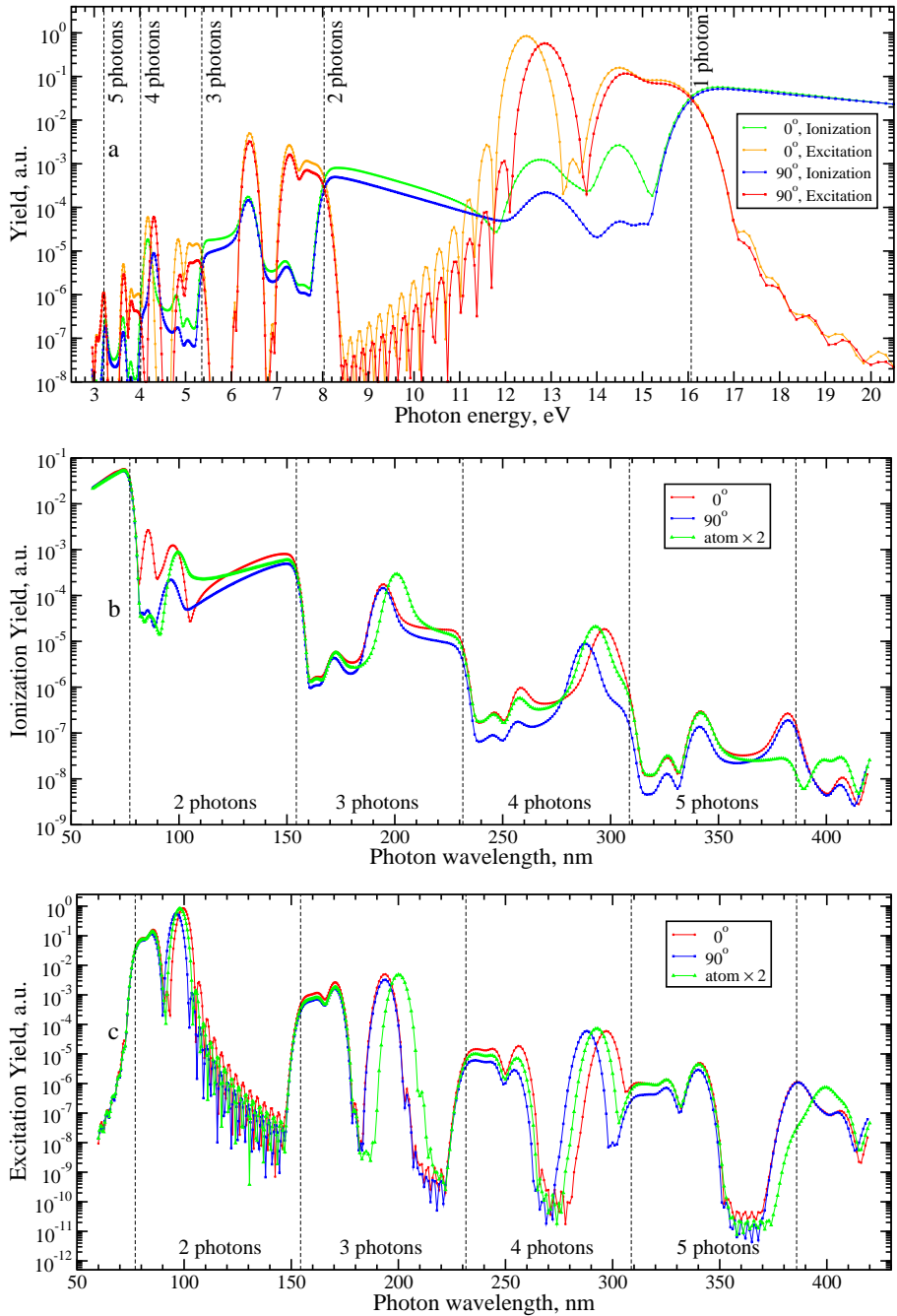}%
\caption{\label{fig:30R1p4} (Color online) 
As Fig.\,\ref{fig:10R1p4}, but for 30-cycle laser 
pulses with peak intensity $5 \cdot 10^{12}$\, W/cm$^2$.} 
\end{center}
\end{figure}
\begin{figure}
\begin{center}
\includegraphics[width=12.0cm]{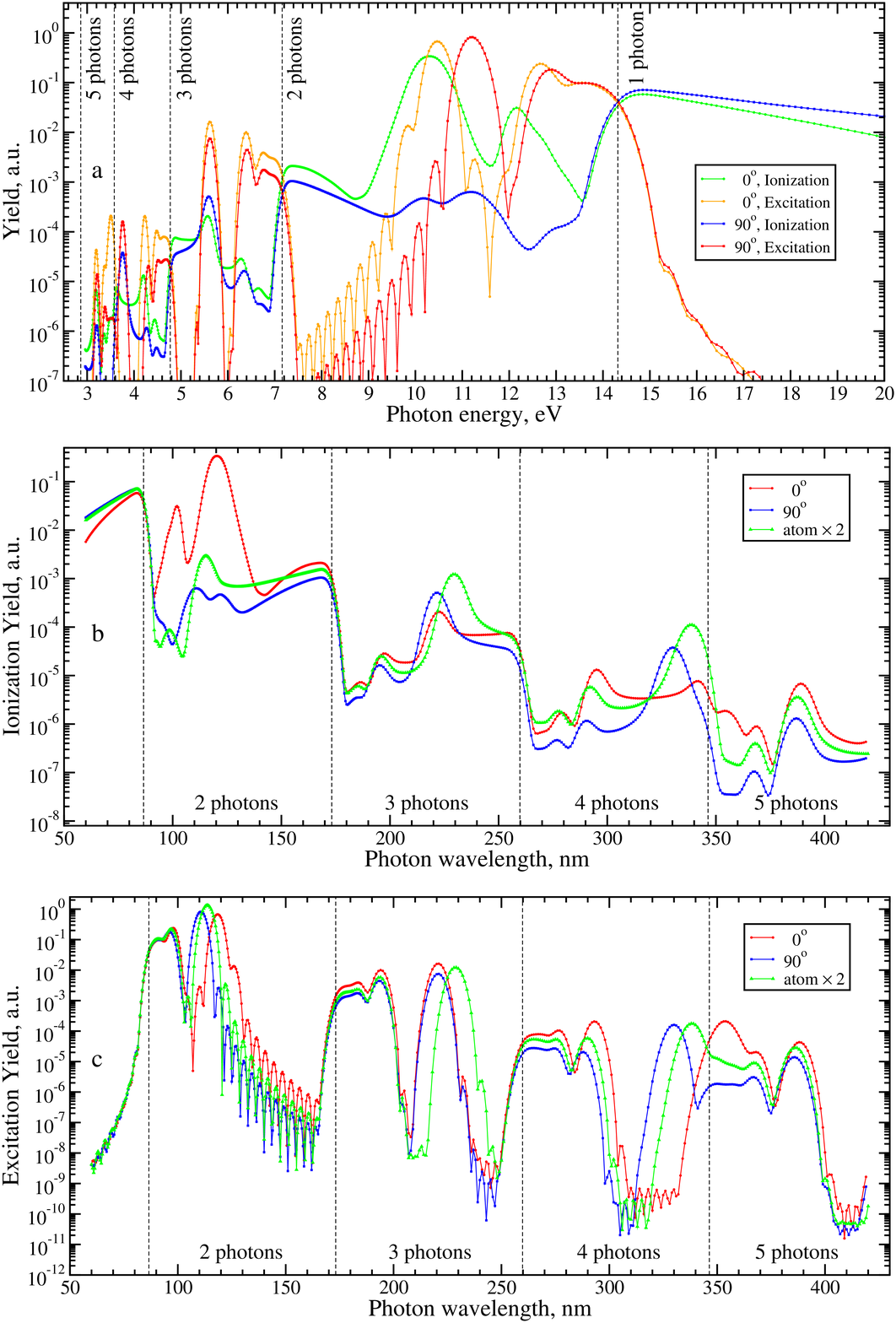}%
\caption{\label{fig:30R2p0} (Color online) As Fig.\,\ref{fig:30R1p4}, but for 
internuclear distance $R = 2.0\,a_0$.} 
\end{center}
\end{figure}
In order to substantiate these arguments even more and to provide 
a more detailed understanding of the influence of the effect of the 
pulse length on the photon-energy resolved multi-photon spectra, 
a second series of calculations was performed in which the pulse length 
was extended to 30 cycles. The results are shown in Fig.\,\ref{fig:30R1p4}. 
Clearly, the ionisation and excitation spectra become much more 
structured. For example, in the already discussed energy range 
from 11 to 16\,eV one sees now in both the excitation and ionisation 
yields well separated peaks due to the (1+1)-REMPI process via the 
B state. Also the B' resonance is clearly visible in the ionisation 
signal and (though less pronounced) also in the excitation yield. 
Furthermore, the cut-off due to the closing of the one-photon ionisation 
channel is much sharper for the 30-cycle pulse --- in direct accordance 
with the smaller band width with which the threshold is convoluted. 

One notices also that the 30-cycle pulse leads to a more pronounced 
REMPI signal, especially for the B state. This reflects the fact that 
the REMPI process needs some time to occur, since the pulse has to 
be sufficiently long to first populate and then ionise the resonant 
intermediate state. Interestingly, the B' state leads to a 
larger ionisation yield than the B state, although the latter is 
much more populated. Evidently, the B' state is more easily ionised 
with the given photon frequency than the B state with the photon 
energy required for its resonant excitation. All these findings for 
different pulse lengths underline the need for full time-dependent 
calculations, if ultrashort pulses are considered.

\begin{figure}
\begin{center}
\includegraphics[width=10.0cm]{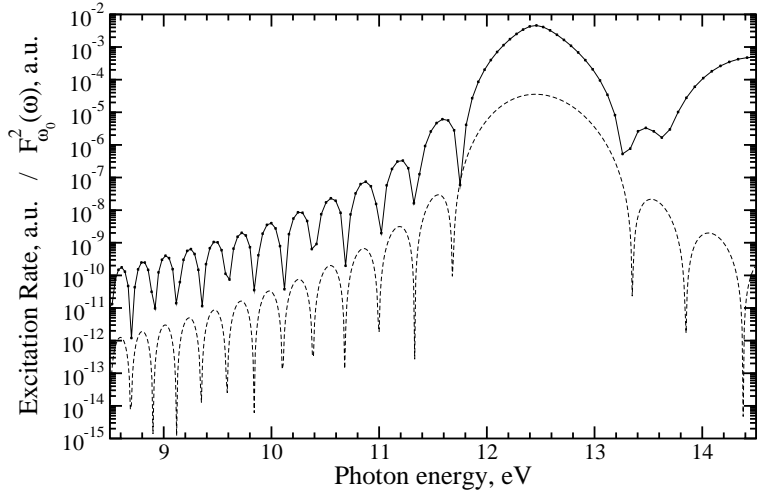}%
\caption{\label{fig:Four} 
Excitation rate (solid line, extracted from the excitation 
yield using equation (\ref{ExcRate})) for 
an H$_2$ molecule with fixed internuclear distance $R = 1.4\,a_0$, 
a 30-cycle linear-polarised laser pulse with 
peak intensity $5 \cdot 10^{12}$\, W/cm$^2$, and a parallel orientation 
of the molecular axis with respect to the field.
It is compared with the square of the Fourier component,
$F^2_{\omega_0}(\omega)$ (dashed line, defined in Eq.\,(\ref{FourCom})),
where $\omega_0$ is the transition frequency from state X to state B.} 
\end{center}
\end{figure}
The excitation spectra in Figs.\,\ref{fig:30R1p4}\,a and c show in the 
energy range from about 8.5 to 12\,eV (100 to 150\,nm) a very pronounced 
oscillatory structure that is absent for the 10-cycle pulses. The origin 
of this feature is the chosen pulse profile ($\cos^2$-type envelope 
function). Its relatively sharp turn-on and -off (compared to a Gaussian 
pulse) which is helpful for 
numerical calculations, since the pulse has a well defined duration 
leading to a clear interval for the time integration, leads to a 
corresponding oscillatory behaviour in the energy domain and thus in the 
convolution function. A semi-quantitative model for this phenomenon 
may be obtained in the following way. For the given laser intensity 
and the photon energy required for a resonant transition to the 
B state (12.46\,eV according to Table \ref{tab:data}) 1st-order 
perturbation theory should provide a reasonable approximation 
for the excitation process. Within lowest-order perturbation theory 
the rate (and thus for a sufficiently long pulse also the yield) 
for an $N$-photon transition is proportional to $I^N$ where 
$I\propto |F|^2$ is the intensity and $F$ the corresponding electric field 
strength. The Fourier component of an $N_c$-cycle laser pulse with 
carrier frequency $\omega$ (and thus the total pulse duration 
$T_p = 2 \pi N_c/\omega$) at a particular frequency $\omega_0$ can be 
obtained from the full time-dependent electric field $F(\omega,t)$ 
by a Fourier analysis (inverse Fourier transform), 
\begin{equation}
 F_{\omega_0}(\omega) = \frac{2}{T_p} \int^{T_p/2}_{-T_p/2} {\rm d}\, t\, 
                        F(\omega,t)\, \cos(\omega_0 t) \quad .
\label{FourCom}
\end{equation}
In Eq.\,\ref{FourCom} it was used that the $\cos^2$ laser pulses 
(with a carrier-envelope phase set to zero) considered in this work 
lead to an even function $F(\omega,t)$. Therfore, the imaginary 
part of the Fourier transform vanishes. 
 
Within the perturbative model outlined above the resonant 1-photon 
transition rate to the B-state is convoluted with $|F_{\omega_0}(\omega)|^2$ 
where $\omega_0$ is equal to the transition frequency from 
the ground state to state B ($12.46\,$eV). Fig.\,\ref{fig:Four} shows 
this convolution function together with the total excitation rate 
obtained from the numerical solution of the TDSE. In order to obtain a 
rate from the calculated excitation yield $Y_{\rm exc}$ the approximate 
relation 
\begin{equation}
 R_{\rm exc} \approx - \ln( 1 - Y_{\rm exc}) / T_p
\label{ExcRate}
\end{equation}
was used. Clearly, the simple model explains very well the observed 
oscillatory structure. 

Although all sharp features (REMPI peaks or ionisation thresholds) 
are, of course, convoluted with the same spectral function, the 
result is only visible if it stems from a very pronounced, 
i.\,e.\ intense and well isolated, signal like the REMPI peak due 
to the B state. Owing to its very low relative intensity, this structure 
is in other cases very easily hidden in the background. For example, 
the oscillations are symmetric around the position of the B-state 
resonance, but on the high-energy side all but the first side band 
are completely covered by the excitation yield due to direct transitions 
to the B' and higher lying Rydberg states. Thus these oscillations are 
almost only visible in the low-energy side where due to energy conservation 
no excited state (B is the lowest-lying one) can directly be populated, 
while the B state can be reached due to the spectral width of the 
$\cos^2$-shaped pulse.  
 
The features and trends discussed above mainly for the 2-photon 
ionisation regime (including the range of (1+1)-REMPI processes 
and the 1-photon ionisation threshold) occur also for the other 
energy ranges considered, but there they are less clearly visible. 
A better visibility and linearity of the spectrum with respect to 
the number of photons involved (in a perturbative picture) is obtained, 
if the yields are shown as a function of the photon wavelength as 
is done in the middle (lower) panels of Figs.\,\ref{fig:10R1p4} and 
\ref{fig:30R1p4} for the ionisation (excitation) yields. In all 
considered regimes up to five photons the longer pulses lead to 
better resolved REMPI peaks and sharper cut-offs at the multi-photon 
thresholds. 

Figs.\,\ref{fig:10R1p4} and \ref{fig:30R1p4} 
show also the results for perpendicular orientation (and $R=1.40\,a_0$). 
Clearly, the overall spectra look very similar for parallel and 
perpendicular orientation when a logarithmic scale is used. In the 
case of 10-cycle pulses both parallel and perpendicular excitation 
yields show clearly one-photon absorption to the electronic excited 
states in the energy window of 11 to 16\,eV. It is, of course, 
different electronic states that are excited in the two cases, since 
the dipole selection rule leads to the excitation of $^1\Sigma_g$ 
states for parallel and of $^1\Pi_u$ states for perpendicular 
orientation. Noteworthy, the corresponding ionisation yield 
shows almost no trace of enhancement in this energy window, 
although the excitation to the lowest lying C state is only a 
factor 2 to 3 lower than the one to the B state, and the 
excitation to the higher lying states is practically identical.

For the 30-cycle pulses REMPI peaks become clearly visible in this 
energy window, but they are much weaker than the corresponding 
peaks obtained for parallel orientation, despite the again very 
similar excitation yields for the corresponding resonant states. 
This indicates that the one-photon ionisation of the $\Pi_u$ 
states (C, C', etc.) with linear-polarised laser light 
perpendicular to the molecular axis possesses a much lower probability 
than the corresponding process for the $\Sigma_u$ states using parallel 
orientation. Such a pronounced difference between the 
resonantly-enhanced ionisation yield and the excitation yield 
of the resonant state is, however, only observed for the 
(1+1)-REMPI processes. Already in the 3-photon regime in which 
(2+1)-REMPI processes occur the ionisation enhancement is to a 
good approximation proportional to the excitation yield, if 
REMPI-processes in parallel and perpendicular direction are 
compared. Within the same orientation (either parallel or 
perpendicular) the REMPI enhancement is now larger for the 
dominant excitation peak (at lower energy) than for the one 
at higher energy. 

\begin{figure}
\begin{center}
\includegraphics[width=13.0cm]{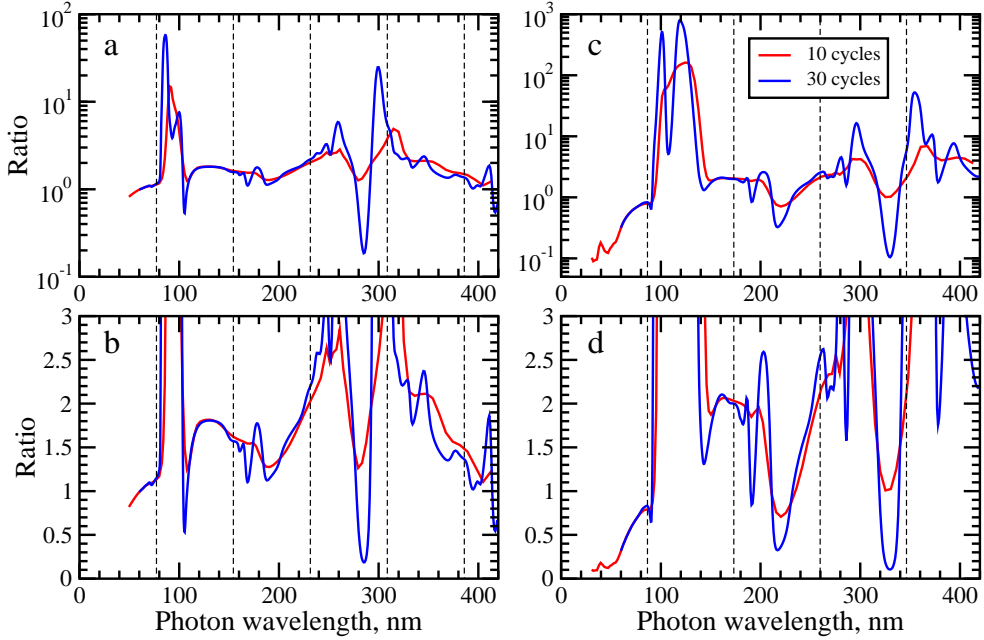}%
\caption{\label{fig:Ratios} (Color online) 
Ratio of the ionisation yields for parallel 
and perpendicular orientation for the internuclear distances 
$R = 1.4\,a_0$ (a,b) and $R = 2.0\,a_0$ (c,d) on either a logarithmic 
(a,c) or a linear (b,d) scale. Shown are the results 
for 10-cycle laser pulses with peak intensity $10^{13}$\,W/cm$^2$ 
(red) and for 30-cycle pulses with peak intensity 
$5 \cdot 10^{12}$\, W/cm$^2$ (blue).} 
\end{center}
\end{figure}
In fact, as already mentioned, if the region containing (1+1)-REMPI 
processes is ignored, the parallel and perpendicular ionisation yields 
are very similar. This is even the case for the excitation yields 
that are only slightly shifted with respect to each other due to the 
energy differences of the bound states with different angular 
momentum $\Lambda$. A closer look reveals that apart from the 
the 1-photon regime where parallel and perpendicular yields 
are almost identical (except the ionisation yields at very high 
photon energies where perpendicular orientation gives higher 
yields) the ionisation and excitation yields are rather uniformly 
larger for parallel than for perpendicular orientation. The 
ratio between parallel and perpendicular ionisation yields is 
shown in Fig.\,\ref{fig:Ratios}\,a for the 10- and 30-cycle 
pulses (and $R=1.40\,a_0$). Despite some sharp resonant features 
that are better resolved for the longer pulses the ratio is 
rather independent of the pulse duration and varies in between 
about 1.5 and 2.5 in the whole 2- to 5-photon regime. This 
overall finding is in reasonable agreement with the corresponding 
results of 2- and 3-photon lowest-order perturbation theory in 
\cite{sfm:apal02}. 

In a recent time-dependent SAE 
study \cite{sfm:niko07} a rather large deviation to the LOPT 
result in \cite{sfm:apal02} was found for a photon energy of 
10.85\,eV ($\approx 115\,$nm). The ratios parallel to perpendicular 
were about 20 \cite{sfm:niko07} or 1.8 \cite{sfm:apal02}. As is 
clearly visible from Fig.\,\ref{fig:Ratios}, the ratio varies 
rather drastically around this wavelength. The present work 
gives a ratio of 1.6 at this photon frequency. However, it 
increases to about 1.8 already at a wavelength of about 120\,nm. 
On the other hand, in between about 115 and 105\,nm the ratio 
drops down to 0.5 before it sharply increases at about 100\,nm. 
As can be especially seen from Fig.\,\ref{fig:30R1p4}\,a, the 
reason for this pronounced variation of the ratio is  
the fact that there is a minimum in the ionisation yield for  
parallel orientation close to this photon energy which may be 
caused by an interference with the REMPI peak caused by the 
B state. This minimum together with the REMPI peak that occurs 
for slightly lower energies than the one due to the C state 
(perpendicular orientation) leads to the sharp minimum followed 
by a sharp maximum, if the photon wavelength approaches the 
region around 125\,nm from above. Clearly, depending on the 
accuracy of the transition energy to the B state provided by 
a particular electronic-structure model very different results 
may be obtained for the ratio between parallel and perpendicular 
orientation at a wavelength around 115\,nm. 

Similarly, it is difficult to compare to the other recent work 
discussing orientational dependence of the ionisation behaviour of 
H$_2$ within TD-DFT \cite{sfm:uhlm06}. The present results agree 
to that work in the sense that no pronounced (order-of-magnitude) 
difference between the parallel and perpendicular orientation is found. 
However, the results in \cite{sfm:uhlm06} are either close to 
saturation of single ionisation (Fig.\,4) or the corresponding 
value at $R=1.4\,a_0$ is barely readable from the graph (Fig.\,5). 
Furthermore, at the considered wavelength of 266\,nm there is 
some structure due to REMPI peaks. Therefore, the exact result 
depends again critically on the corresponding excitation 
energies (and spectral width of the laser pulse). Noteworthy, 
the present calculation confirms resonant structure at this 
wavelength as is also noted (for this $R$ value) 
in \cite{sfm:uhlm06}.

It is usually assumed that few-photon processes depend heavily 
on the electronic structure, if not too high intensities are considered, 
while many-photon processes depend mostly on the ionisation 
potential and shape of the long-range potential experienced by 
the ejected electron. Accordingly, one would not expect an 
atomic model like the one described in Sec.\,\ref{sec:modelatom} 
to yield ionisation and excitation yields in reasonable 
agreement to a full molecular calculation for few-photon 
processes. However, as is evident from Figs.\,\ref{fig:10R1p4} 
and \ref{fig:30R1p4}, the overall agreement even of the excitation 
yield is in fact surprisingly good. The ionisation yield 
agrees in the 1-photon regime almost perfectly with the 
molecular result for perpendicular orientation. Especially for 
the 30-cycle pulses the agreement continues in the 2-photon 
regime until the second REMPI peak where the atomic model 
agrees quantitatively much better with the parallel result. 
At the threshold between 2- and 3-photon ionisation (at 
about 155\,nm) all three curves agree very well for both pulse 
lengths. Once the parallel and perpendicular results start to 
disagree, the atomic result follows now more closely the molecular 
results for parallel orientation. Starting at about 360\,nm 
the atomic results disagree with the molecular ones that 
agree at these wavelengths rather well with each other. 
A very similar result is also found for the excitation 
yields. This confirms that even the position (and dipole 
moments) of the excited states are very well approximated 
by the atomic model proposed in this work. 

Finally, it is interesting to investigate whether the findings 
reported so far apply only to the case when the internuclear 
distance is relatively small and thus the ground-state 
electron density is rather isotropic. A second series of 
calculations was thus performed in which $R=2.0\,a_0$ was 
used. The results are given in Figs.\,\ref{fig:10R2p0} and 
\ref{fig:30R2p0} as well as in the right panel of 
Fig.\,\ref{fig:Ratios}. 
This choice of $R$ is motivated by the fact that the ground 
vibrational wavefunction of H$_2$ extends to about this 
distance and it is at the same time the equilibrium distance of 
the H$_2^+$ ion created in the ionisation process. 

Most evidently, the 1- and 2-photon ionisation yields disagree 
for the two orientations much more than for $R=1.40\,a_0$. Only 
at the threshold between the two regimes reasonable to good 
agreement is found. While the 1-photon yield for parallel 
orientation lies below the perpendicular one, this changes 
drastically in the 2-photon regime. The latter difference seems to 
be mainly due to the now even much more pronounced REMPI peaks for 
parallel orientation. Noteworthy, the excitation yields are 
very similar for the two orientations, besides the energy 
shifts due to different excitation energies. In the regimes of 
3- to 5-photon ionisation the qualitative difference between 
the two orientations is less pronounced than in the 2-photon 
case. Compared to $R=1.4\,a_0$ it is especially the quantitative 
agreement between the two orientations which becomes worse, if 
$R$ increases. This is also seen from the corresponding ratio 
(Fig.\,\ref{fig:Ratios}) which covers a larger range of values 
in this case. It is also interesting to note that a comparison 
of the results for $R=1.4$ and $2.0\,a_0$ shows that there is 
quite good qualitative agreement. The main difference is that 
with increasing $R$ value the excitation energies become smaller. 
Together with a decrease of the vertical ionisation potential 
when going from $R=1.4$ to $2.0\,a_0$ this leads to an effective 
compression of the wavelength ranges for a given $N$-photon 
regime. The, of course, $R$-independent results of the atomic model 
agree not as well with the molecular yields for larger than 
for smaller $R$, as is expected. Nevertheless, the atomic model 
still works reasonably well for a qualitative or even semi-quantitative 
estimate.

\section{Conclusion and outlook}
A previously developed numerical approach for solving the time-dependent
Schr\"odinger equation that describes the correlated motion of both 
electrons of molecular hydrogen exposed to a short
intense laser pulse in the non-relativistic, fixed-nuclei, and
dipole approximation has been extended to consider also a perpendicular
orientation of the molecular axis with respect to the laser field.
The ionisation and excitation yields were calculated for photon
wavelengths of about 40 to 420\,nm covering the complete 2- to 5-photon
regime and extending into the 1- and 6-photon ranges.
Two pulse lengths (10- or 30-cycle pulses) and two internuclear
separations (1.4 and 2.0\,$a_0$) were considered.

Mainly due to the smaller spectral width (but partly also due to the
lower peak intensity), the longer pulses yield much more structured
spectra than the short ones. However, due to the intrinsic time
dynamics of the REMPI process as a consequence of the implied two-step
process (population and depopulation of the resonant state), the different
REMPI signals show a different pulse-length dependence.

It is found that the parallel and perpendicular spectra agree
qualitatively quite well with each other, the main difference
stems from the difference in the positions of the resonant peaks due
to the different symmetries of the corresponding intermediate states.
Even the quantitative differences are not too large in the non-resonant
parts of the spectra. At the equilibrium distance of H$_2$ the ratio
of parallel to perpendicular non-resonant multi-photon ionisation yields 
varies in between
1 and 2.5. It is therefore only weakly anisotropic. For the increased
distance $2.0\,a_0$ this ratio tends especially in the 4- and 5-photon
regimes to slightly larger values up to about a value of 4. The ratios
found in the 2- and 3-photon regimes
agree reasonably well with previous ones obtained with perturbation
theory \cite{sfm:apal02}, but disagree with the much larger value
obtained at a single photon frequency in the 2-photon regime
using a single-active-electron approximation \cite{sfm:niko07}.
However, very large ratios are found in the resonant regimes.
Since the position of the resonances depends on the quality of
the used electronic-structure model, large discrepancies may occur,
if different levels of approximation are used.

An atomic model is proposed that allows to easily tune the
ionisation potential while maintaining asymptotically the
shape of the appropriate long-ranged Coulomb potential. This allows to
more clearly analyse molecular effects that are caused by the
multi-centred character of a molecular electron density.
Surprisingly, even (in fact especially) in the few-photon
regime the single-electron atomic model compares very well
with both the ionisation and excitation yield obtained with the
full molecular two-electron calculation.

Presently, this work is extended to longer wavelengths (especially
the popular 800\,nm of the Ti:sapphire laser) and to the analysis
of differential quantities like above-threshold-ionisation (ATI)
electron spectra. Furthermore, the anisotropy of high-harmonic
emission of H$_2$ will be investigated.

\section*{Acknowledgments}
The authors acknowledge financial support by the {\it Deutsche
Forschungsgemeinschaft} (DFG-Sa\,936/2). AS is grateful to the
{\it Stifterverband f\"ur die Deutsche Wissenschaft} (Program
{\it Forschungsdozenturen}) and the {\it Fonds der Chemischen Industrie}
for financial support.

\bibliographystyle{tMOP}

\end{document}